\documentclass[a4paper,USenglish]{lipics-v2021}

\hideLIPIcs  

\title{Reaching Agreement Among \emph{k} out of \emph{n} Processes}

\author{Gadi Taubenfeld}{Reichman University, [Herzliya 46150], Israel \and \url{https://faculty.runi.ac.il/gadi}}{tgadi@idc.ac.il}{https://orcid.org/0000-0003-3070-5370}{}

\authorrunning{G. Taubenfeld}

\Copyright{Gadi Taubenfeld}

\ccsdesc[500]{Theory of computation~Distributed computing models}
\ccsdesc[500]{Theory of computation~Shared memory algorithms}
\ccsdesc[500]{Theory of computation~Distributed algorithms}

\keywords{full agreement, partial agreement, shared memory, message passing.}

\nolinenumbers 

\EventEditors{John Q. Open and Joan R. Access}
\EventNoEds{2}
\EventLongTitle{42nd Conference on Very Important Topics (CVIT 2016)}
\EventShortTitle{CVIT 2016}
\EventAcronym{CVIT}
\EventYear{2016}
\EventDate{December 24--27, 2016}
\EventLocation{Little Whinging, United Kingdom}
\EventLogo{}
\SeriesVolume{42}
\ArticleNo{23}

\begin{document}

\maketitle

\def\propose{{\sf propose}}
\def\decision{\mathit{decision}}
\newcommand{\abs}[1]{ }

\begin{abstract}
In agreement problems,
each process has an input value and must choose
a decision (output) value.
Given $n\geq 2$ processes
and $m \geq 2$ possible different input values,
we want to design an agreement algorithm that enables as many processes
as possible to decide on the (same)
input value of one of the processes,
in the presence of $t$ crash failures.
Without communication, when each process
simply decides on its input value, at least $\lceil (n-t)/m \rceil$ of the processes are guaranteed to always decide on the same value. Can we do better with communication?
For some cases, for example when $m=2$, even in the presence of a single crash failure,
the answer is negative in a deterministic asynchronous system where
communication is either by using atomic read/write registers or by sending and receiving messages.
The answer is positive in other cases.
\end{abstract}

\maketitle

\section{Introduction}
\label{sec:Intro}
The problem of reaching agreement
is a fundamental coordination problem and is at the
core of many algorithms for fault-tolerant distributed applications.
The problem is to design an algorithm in which \emph{all} the participants
reach a common decision based on their initial opinions.
This problem is a special case of the $(n,k)$-partial agreement problem, introduced and defined below,
in which it is required that at least $k$ of the $n$ participants reach a common decision.
When the exact values of $n$ and $k$ are not important,
we will refer to this problem as the \emph{partial agreement} problem.

The relation of the 
notion of partial agreement with the interesting notions of $X$ agreement \cite{DPPU1988},
almost everywhere agreement
(in which all but a small number of correct participants must choose a common decision value) \cite{DPPU1988},
almost-$t$-resilient agreement (allowing a limited number of correct participants not to terminate)
\cite{Tau2018journal},
and bounded disagreement \cite{CHS2020}, is discussed in details in the related work section.

\subsection{The (\emph{n,k})-partial agreement problem}
The \emph{$t$-resilient} \emph{$(n,k)$-partial agreement} problem is to design an algorithm
for $n$ processes that supports a single operation called $\propose()$, and can tolerate $t$ crash failures.
The operation takes an input parameter, called the \emph{proposed}
value, and returns a result, called the \emph{decided} value.
It is assumed that each of the $n$ processes invokes the propose operation at most once.
The problem requirements are that there exists
a {\em decision value} $v$ such that:
%
\begin{itemize}
\item
\emph{Agreement}:
At most $n-k$ processes may decide on values other than $v$. Thus,
when all the $n$ processes decide, at least $k$ of them decide on (the same value) $v$.
\item
\emph{Weak validity}: $v$ is the input (proposed) value of at least one of the processes.
\item
\emph{$t$-resiliency}: Each process that does not crash eventually decides and terminates,
as long as no more than $t$ processes crash.
\end{itemize}
We notice that the agreement requirement means that, in every execution,
there must exist a value $v$ such that
the number of processes that have decided on $v$ plus the number of processes
that haven't decided yet (possibly crashed) is at least $k$.
The $(n,n)$-partial agreement problem is the familiar consensus (i.e., full agreement)
problem for $n$ processes \cite{PSL1980},
in which all the non-faulty processes must eventually decide on the same value,
which must be a proposed value.
When there are only two (resp.\ more than two) possible input values,
the problem is called the partial binary (resp.\ partial multi-valued) agreement problem.\\
\\
Weak validity only requires that $v$ be a proposed value. A stronger validity requirement is,
\begin{itemize}
\item
\emph{Strong validity}: Every decided value must be a proposed value.
\end{itemize}
The necessary conditions proved in section \ref{sec:multiValued} hold only for
strong validity. 
All the other results in the paper hold for both the weak and strong validity requirements.

%
\subsection{The (\emph{n,k,l})-partial set agreement problem}
The \emph{$t$-resilient} \emph{$(n,k,\ell)$-partial set agreement} problem
captures a weaker form of the $(n,k)$-partial agreement problem in which the
agreement property is weakened.
The problem is to design an algorithm
for $n$ processes that supports a single operation called $\propose()$
and can tolerate $t$ crash failures.
The operation takes an input parameter, called the \emph{proposed}
value, and returns a result, called the \emph{decided} value.
It is assumed that each of the $n$ processes invokes the propose operation at most once.
The requirements of the problem are that there exists
a set of decision values $V$ of size at most $\ell$ such that:
%
\begin{itemize}
\item
\emph{Agreement}:
At most $n-k$ processes may decide on values not in $V$. Thus,
when all the $n$ processes decide, at least $k$ of them decide values in $V$.
\item
\emph{Weak validity}: Each $v\in V$ is the input (proposed) value of at least one of the processes.
\item
\emph{$t$-resiliency}: Each process that does not crash eventually decides and terminates,
as long as no more than $t$ processes crash.
\end{itemize}
%
As before, a stronger
validity
requirement, called
\emph{strong validity}, is that every decided value must be a proposed value.
The $(n,n,1)$-partial set agreement problem is the familiar consensus problem.
The $(n,k,1)$-partial set agreement problem is the $(n,k)$-partial agreement defined earlier.
The $(n,n,\ell)$-partial set agreement problem, with strong validity, is the familiar $\ell$-set agreement problem
for $n$ processes, which is
to find a solution where each process starts with an input value from some domain
and must choose some
process' input as its output, and
all the $n$ processes together may choose no more than $\ell$ distinct output values \cite{Chaudhuri1993}.
\subsection{Motivation}
The first and foremost motivation for this study is related to the
basics of computing, namely, increasing our knowledge of what can (or cannot) be done in
the context of failure-prone distributed systems.
Providing necessary
and sufficient conditions for the solvability of the partial agreement problem helps us determine
the limits of synchronization algorithms and identify
to what extent communication is helpful when solving weak
variants of the fundamental \emph{full} agreement problem.
Furthermore, as was pointed out in \cite{DPPU1988},
in many practical situations, we may be willing to settle
for cooperation between the vast majority of the processes,
which raises the question of when this is possible.

Another inspiration for this work is related to biology.
The partial agreement problem arises in biological systems where there is a predefined threshold,
and it is only required that the number of participants that reach agreement exceeds the threshold
for a specific action to take place.
A well-known and extensively studied example is \emph{quorum sensing}.
Many species of bacteria use quorum sensing
to coordinate gene expression according to the density of their local population.
Quorum sensing is triggered to begin when the number of bacteria, that sense that
a sufficient number of bacteria are present, reaches a certain threshold.
Quorum sensing allows bacteria to synchronize and, by doing so,
enables them to successfully infect and cause disease in plants, animals,
and humans \cite{BB2019,FWG1994}.
Also, several groups of social insects, like ants and honey bees, have been shown to use quorum sensing
in a process that resembles collective decision-making \cite{Pratt2005,SV2004}.
%
%
\subsection{Models of computation}
Our model of computation consists of a collection
of $n$ deterministic processes.
Each process has a unique identifier.
We denote by $t$ the maximum number of processes that may fail.
The only type of failure considered in this paper is a process \emph{crash} failure.
A crash is a premature halt. Thus, until a process possibly crashes,
it behaves correctly by reliably executing its code.
We consider the following shared memory (SM) and message passing (MP) 
models.
%
\begin{enumerate}
\item
\emph{The asynchronous RW model.}
There is no assumption on the relative speeds of the processes.
Processes communicate by atomically reading and writing shared
registers. A register that can be written and read by any process is a multi-writer multi-reader
(MWMR) register.  If a register can be written by a single (predefined)
process and read by all, it is a single-writer multi-reader  (SWMR)
register.
\item
\emph{The asynchronous MP model.}
There is no assumption on the relative speeds of the processes.
Processes communicate by sending and receiving messages.
There is no assumption on the speed of the messages.
\item
\emph{The synchronous MP model.}
It is assumed that the processes communicate in ``rounds'' of communications.
At the beginning of a round, each process may send messages to other processes,
and all the messages sent during a round arrive at their destinations
by the end of this round. Processes start each round at the same time.
That is, a process may start participating in a round only
when all other processes have finished participating in the previous round.
\item
\emph{The asynchronous SM(g) model.}
There is no assumption on the relative speeds of the processes.
Processes communicate by (1) atomically reading and writing shared
registers, and (2) using full agreement objects for $g$ processes
that can tolerate any number of failures -- also called wait-free
full agreement objects for $g$ processes.
\end{enumerate}
\subsection{Known results}
We will use the following known results:
\begin{enumerate}
\item
There is no solution for the $(n,n)$-partial agreement problem
(i.e., consensus problem)
for $n\geq 2$ processes and $m\geq 2$ input (proposed) values
that can tolerate a single crash failure in an asynchronous system where communication is
done either by sending messages or by reading and writing atomic registers \cite{FLP85,LA87}.
\item
For any $\ell \geq 1$,
there is no solution for the $(n,n,\ell)$-partial set agreement problem, assuming strong validity
(i.e., $\ell$-set agreement problem),
for $n\geq \ell +1$ processes and $m\geq \ell +1$ input values
that can tolerate $\ell$ crash failures in an asynchronous system where communication is
done either by sending messages or by reading and writing atomic registers \cite{BG93,HS99,SZ2000}.
\item
In a synchronous message-passing system in which up to $1\leq t\leq n-2$ processes may crash,
every full agreement algorithm  requires at least $t + 1$ rounds~\cite{AT1999,DS1983},
and there exists a full agreement algorithm with $t+1$ rounds~\cite{AW98}.
When $t=n-1$, $t$ rounds are necessary and sufficient.
\item
There is no solution for the $(n,n)$-partial agreement problem
for $n\geq t+1$ processes
that can tolerate  $t$ crash failures in an asynchronous system using
atomic registers and  wait-free full agreement objects for $t$ processes \cite{Her91}.
\end{enumerate}
\newpage
\subsection{Content of the article}
Let $n$ be the number of processes,
$m$ the number of possible different input values,
$t$ an upper bound on the number of crash failures, and
$g$ the size (\# of processes) of the full agreement objects
when assuming the $SM(g)$ model.
In all the results, unless stated otherwise, it is assumed that $n\geq 2$.
Given two positive integers $a$ and $b$, the notation $a \bmod b$ (i.e., $a$ modulo $b$)
is used for the remainder of the division of $a$ by $b$.
Table \ref{table:global-view} summarizes
the main results presented in this article
regarding the solvability and complexity of the $(n,k)$-partial agreement problem.
%
\begin{table}[!h]
\begin{center}
\renewcommand{\baselinestretch}{1}
\begin{tabular}{|l||l|l|l|l|l|l|}
\hline
\multicolumn{7}{|c|}{\textbf{Necessary and sufficient conditions for solving the $(n,k)$-partial agreement problem}}\\
\multicolumn{7}{|c|}{The necessary conditions proved in section \ref{sec:multiValued} (i.e., R4 \& R5) hold only assuming
strong validity.}\\
\multicolumn{7}{|c|}{All the other results hold for both the weak and strong validity conditions.}\\
\hline
Res-\ & Model & Values & Necessary & Sufficient & Comm. & Sec-\\
ult        &       &  & condition   & condition  &helps? &tion\\
\hline
\hline
R1 & Asynchronous & $m=2$ & $k \leq \lceil n/2 \rceil$
& $k \leq \lceil n/2 \rceil$ & No & \ref{BinaryAgreement}\\
&RW + MP&$t\geq 1$ &Theorem~\ref{Thm:BinaryAgreement}&Theorem~\ref{Thm:BinaryAgreement}&& \\
\hline
R2 & Asynchronous & $m\geq 2$ & $k \leq \lceil n/2 \rceil$
& $k \leq \lceil n/2 \rceil$ & Yes, when & \ref{BinaryAgreement}\\
&RW + MP&$t= 1$ &Corollary~\ref{coro:singleFailure}&Corollary~\ref{coro:singleFailure}& $m > 2$& \\
\hline
R3 & Asynchronous & $m\geq 2$ & $k \leq \lceil n/2 \rceil$
&        & & \ref{BinaryAgreement}\\
&RW + MP&$t\geq 1$ &Corollary~\ref{coro:BinaryAgreementPlus}& & & \\
\hline
\hline
R4 & Asynchronous & $m\geq 2$ & $k \leq $
& $k \leq $ & No, when & \ref{sec:multiValued}\\
&RW + MP&$t\geq 1$ &$ n/\min(m,t+1)$&$n/\min(m,t+1)$ & $m\leq t+1$& \\
    &    &$\min(m,t+1)$&   &   &   &  \\
\multicolumn{1}{|l||}{} &
\multicolumn{1}{|l|}{} &
\multicolumn{1}{|l|}{divides $n$} &
\multicolumn{1}{|l|}{Corollary~\ref{coro:remainder}} &
\multicolumn{1}{|l|}{Corollary~\ref{coro:remainder}} &
\multicolumn{1}{|l|}{Yes, if not} &
\multicolumn{1}{|l|}{} \\
\hline
R5 & Asynchronous & $m\geq 2$ & $k \leq $
& $k \leq $ & No, when & \ref{sec:multiValued}\\
&RW + MP&$t\geq 1$ &$\lfloor n/ \min(m,t+1) \rfloor +$&$\lceil n/ \min(m,t+1) \rceil$  & $m\leq t+1$& \\
& & &$n\bmod \min(m,t+1)$& &  & \\
& & &Theorem~\ref{thm:MultivalueAgreement}&Theorem~\ref{thm:MultivalueAgreement}& Yes, if not & \\ 
\hline
\hline
R6 & Synchronous & $m \geq 2$ & $t$ rounds are necessary
&    & & \ref{sec:Synchronous} \\
& MP&$t\geq 1$ & for every & & & \\
    &    &    & $k \geq \lceil (n+t+1)/2 \rceil$ &   &   &  \\
& & &Theorem~\ref{thm:lowerBound}& & & \\
\hline
R7 & Synchronous & $m\geq 2$ & & $\lfloor t/\ell \rfloor +1$ rounds & Yes & \ref{sec:Synchronous} \\
& MP&$t\geq 1$ &   &are sufficient for & & \\
& &$\ell \geq 1$ & &$k \leq \lceil n/\ell \rceil$& & \\
& & & &Theorem~\ref{thm:upperBound}& & \\
\hline
\hline
R8 & Asynchronous & $m\geq 2$ & $k \leq \lceil (n+t-1)/2 \rceil$
& & & \ref{sec:shrongSM}\\
&SM($g$)&$n>t\geq 1$ &  &  &  & \\
& &$g = t$ &Theorem~\ref{Thm:strongObjects}& & & \\
\hline
R9 & Asynchronous & $m\geq 2$ &
& $k \leq \max $ & Yes & \ref{sec:shrongSM}\\
&SM($g$)&$t\geq 1$ & & $(\lceil n/ \min(m,t+1),$ & & \\
& & $g\geq 1$ & & $g, 3\lfloor \min(\lfloor n/2 \rfloor ,g)/2 \rfloor )$ &  & \\
& & & &Theorems~\ref{thm:alg:StrongSM} \& \ref{thm:MultivalueAgreement} (R5) & & \\
\hline
R10 & Asynchronous & $m\geq 2$ & $k \leq 3n/4$
&$k \leq 3n/4$ &Yes & \ref{sec:shrongSM}\\
&SM($g$)&$g=t=n/2$ &  &  &  & \\
& &4 divides $n$ &Corollary~\ref{coro:strongObjects}&Corollary~\ref{coro:strongObjects}& & \\
\hline
\end{tabular}
\end{center}
\caption{Summary of the results}
\label{table:global-view}
\vspace{-0.3in}
\end{table}
\newline
A few remarks.
\begin{enumerate}
\item
It follows from R3
that in the presence of failures, the best we can hope
for is to solve $(n,k)$-partial agreement for $k = \lceil n/2 \rceil$.
This can be achieved when either $m=2$ (R1) or $t=1$ (R2).
\item
It is interesting to note that for proving R1--3, it suffices
to use the (above)
known impossibility result \#1.
In contrast, for proving R5, there is a need to use the (much stronger)
known impossibility result \#2.
\item
In proving the necessary condition of R5 (and of R4), strong validity is assumed.
Results R1--3 hold for both weak and strong validity.
\item
R4, which follows from R5, provides a tight bound for partial multi-valued
agreement in the special case where $\min(m,t+1)$ divides $n$.
\item
When looking at the round (time) complexity of synchronous partial agreement,
R6 shows that in many cases (i.e., when $\lceil (n+t+1)/2 \rceil \leq k < n$)
we might be able to save only one round, compared to the solvability of full agreement for which
$t+1$ rounds are necessary and sufficient.
\item
R7 shows that in some cases, it is possible to significantly
reduce the number of rounds,
compared to the solvability of full agreement for which
$t+1$ rounds are necessary and sufficient.
R7 follows easily
from a known result regarding the number of rounds sufficient
for solving the set agreement problem \cite{AW98,CHLT2000}.
\item
It follows from R8 and R9 that for the $SM(n/2)$ model
(i.e., when consensus objects for $n/2$ processes are available),
the bound is tight when $n=t/2$ and $n$ is divisible by 4 (R10).
\end{enumerate}

\section{Asynchronous Partial Agreement: The Binary Case with Implications}
\label{BinaryAgreement}
Let $n\geq 2$ be the number of processes,
$m \geq 2$ the number of possible different input values, and
$t\geq 1$ an upper bound on the number of crash failures.
We show that in the presence of failures, the best we can hope
for is to solve $(n,k)$-partial agreement for $k = \lceil n/2 \rceil$
(Corollary) \ref{coro:BinaryAgreementPlus}).
This bound is tight
when either $m=2$ (Theorem~\ref{Thm:BinaryAgreement}) or $t=1$ (Corollary~\ref{coro:singleFailure}).
We first show that for binary partial agreement, in the presence of failures,
at most $\lceil n/2 \rceil$ processes are
guaranteed to decide on the same value and that this can be achieved
without communication.
All the results in Section~\ref{BinaryAgreement} hold under both the weak and the strong validity requirements.

\emph{Computational model.}
The results presented in Section~\ref{BinaryAgreement} and Section~\ref{sec:multiValued}
hold for a shared memory model that supports atomic read/write registers and
a message passing model that supports send and receive messages.
The necessary conditions (impossibility results) in these
sections will be proven for the shared memory model.
However, we observe that a shared memory system that supports
atomic registers can simulate a message passing system that
supports send, receive and even broadcast operations.
Hence the necessary conditions (impossibility results)
proved for the shared memory model in Section~\ref{BinaryAgreement} and Section~\ref{sec:multiValued}
also hold for such a message passing system.
The simulation is as follows.
With each process $p$ we associate an unbounded array of shared registers
which all processes can read from, but only $p$ can write into.
To simulate a broadcast (or sending) of a message, $p$ writes
to the next unused register in its associated array.
When $p$ has to receive a message, it reads the new messages from each process.

\begin{theorem}
\label{Thm:BinaryAgreement}
For $n\geq 2$, $m=2$ and $t\geq 1$,
there exists an $(n,k)$-partial agreement algorithm that can tolerate $t$ crash failures
if and only if $k \leq \lceil n/2 \rceil$.
Furthermore, for every $k \leq \lceil n/2\rceil$, there exists such an algorithm in which the processes do not need to communicate.
\end{theorem}
Informally, the essence of the proof is in
showing that an $(n,\lceil n/2 \rceil+1)$-partial agreement algorithm (object)
has the same computational power as an $(n,n)$-partial agreement algorithm,
in the presence of a single failure.
Thus, since it is impossible to solve $(n,n)$-partial agreement,
it is also impossible to solve $(n,\lceil n/2 \rceil+1)$-partial agreement.
In the proof, the known result \#1,
regarding the impossibility of solving $(n,n)$-partial agreement
in the presence of a single faulty process, is used.

\proof
Since, in an asynchronous system,
a crashed process cannot be distinguished from a very slow process,
the agreement requirement is equivalent to (i.e., can be simplified as follows):
``When all the $n$ processes decide, at least $k$ of them decide on the same value.''
Without communication, when each process
simply decides on its input value, at least $\lceil n/2 \rceil$ of the processes are guaranteed to decide on the same value
in runs where all the $n$ processes decide (i.e., in fault-free runs).
Obviously, since there is no communication,
this simple algorithm satisfies both the weak and strong validity requirements for any number $t\geq 1$ of failures.
This completes the proof of the \emph{if} direction.

To prove the \emph{only if} direction, we assume to the contrary that
there exists an $(n,k)$-partial agreement algorithm where $k=\lceil n/2 \rceil +1$, called $A$, that can tolerate $1$ crash failure,
and shows that this assumption leads to a contradiction.
Obviously, proving the result for $t=1$ and $k=\lceil n/2 \rceil +1$ implies the same result for $t\geq 1$
and $k=\lceil n/2 \rceil +1$.

By definition, in any (fault-free) run of $A$ in which all the $n$ processes decide,
there must exist a (proposed) value $v$ such that
the number of processes that decide on $v$ minus the number of processes
that decide on any other possible value is at least two
(two when $n$ is even, and three when $n$ is odd).
Moreover, in any run in which \emph{exactly} one process fails,
and all the other processes decide,
there must exist a (proposed) value $v$ such that
the number of processes that decide on $v$ minus the number of processes
that decide on any other possible value is at least one
(one when $n$ is even, and two when $n$ is odd).
Thus, in any run of $A$ in which \emph{at most} one process fails, there is a
(proposed) value $v$ such that a strict majority (i.e., more than half) of the processes decide on $v$.

We use $A$ to construct an $(n,n)$-partial agreement algorithm that can tolerate a single
crash failure, called $B$, as follows:
$B$ works in two (asynchronous) phases of computation:%
\footnote{
There are no synchrony assumptions whatsoever.
A process that finishes phase one,
immediately starts participating in phase two.}
\begin{enumerate}
\item
Phase one: Each process $p$ participates in $A$ and decides on some value denoted $\decision_p(A)$.
\item
Phase two:
Each process $p$ owns a single-writer register, and
initially writes $\decision_p(A)$ in a single-writer register.
Then $p$ repeatedly
reads the single-writer registers of the other processes
until it learns the decision values from the first phase of all
the other processes except maybe one of them (since one process may fail).
\end{enumerate}
As explained above, in the $n-1$ decision values
from phase one that $p$ knows about (including its own value),
there must be one (proposed) value $v$ that was decided upon by more than half of the processes.
So, at the end of phase two, $p$ decides on that value $v$, and terminates.
This completes the description of algorithm $B$.
We prove that when $t= 1$, in $B$ all the non-faulty processes decide on the same value $v$.
Consider two possible cases:
\begin{enumerate}
\item
All the $n$ processes succeed in writing their decision values from phase one into their
single-writer registers. In such a case, as explained above,
there must exist a value $v$ such that (in phase one)
the number of processes that decided on $v$ minus the number of processes
that decided on any other possible value is at least two.
Thus, in any subset of size $n-1$ of these $n$ values (that some process may know
about at the end of phase one) $v$ is the majority value.
\item
Some process failed, and
only $n-1$ processes succeeded in writing their decision values from phase one into their
single-writer registers. In such a case, as explained above,
there must exist a value $v$ such that (in phase one)
the number of processes that decided on $v$ minus the number of processes
that decided on any other possible value is at least one.
Thus, since all the non-faulty processes see (at the end of phase two)
the same subset of size $n-1$, they will all decide on the same (proposed) value $v$.
\end{enumerate}
Thus, $B$ is an $(n,n)$-partial agreement algorithm that can tolerate one faulty process,
violating the known result \#1 (as stated in the introduction),
regarding the impossibility of solving $(n,n)$-partial agreement
in the presence of a single faulty process \cite{FLP85,LA87}.
\qed

\begin{corollary}
\label{coro:BinaryAgreementPlus}
For $n\geq 2$, $m\geq 2$, and $t\geq 1$,
there exists an $(n,k)$-partial agreement algorithm that can tolerate $t$ crash failures
only if $k \leq \lceil n/2 \rceil$.
\end{corollary}

\proof
By definition,
any algorithm that solves the $(n,k)$-partial agreement problem
in the presence of (up to) $t$ failures when the maximum number
of possible input values is $m$, must also solve
the $(n,k)$-partial agreement problem
in the presence of $t$ failures when the maximum number
of possible input values is strictly less than $m$.
The result follows.
\qed

\begin{corollary}
\label{coro:singleFailure}
For $n\geq2$, $m\geq 2$, and $t=1$,
there exists an $(n,k)$-partial agreement algorithm that can tolerate a single crash failure
if and only if $k \leq \lceil n/2 \rceil$.
\end{corollary}

\proof
The only if direction follows immediately from Corollary~\ref{coro:BinaryAgreementPlus}.
For proving the if direction, consider the following
$(n,\lceil n/2) \rceil)$-partial agreement algorithm that can tolerate a single failure.
Each process $p$ writes its input value into a single writer register
(resp.\ sends its input to everybody)
and continuously reads the single-writer registers of the other processes
(resp.\ waits to receive messages from the other processes)
until it knows the inputs of $n-1$ processes, including itself.
Since there is at most one failure, this procedure will always terminate.
Then, $p$ decides on the maximum input value it knows about.
This reduces the number of decision values to at most 2.
Hence, this algorithm  solves the $(n,\lceil n/2\rceil)$-partial agreement problem.
\qed

\section{Asynchronous Partial Agreement: The Multi-valued Case}
\label{sec:multiValued}
We provide necessary and sufficient conditions for the
solvability of multi-valued partial agreement.
These conditions also indicate when communication might help.
The sufficient conditions hold under both the weak and the strong validity requirements;
the necessary conditions hold only under the strong validity requirement.
Given two positive integers $a$ and $b$, the notation $a \bmod b$ (i.e., $a$ modulo $b$)
is used for the remainder of the division of $a$ by $b$.

\begin{theorem}
\label{thm:MultivalueAgreement}
For $n\geq2$, $m\geq 2$ and $t\geq 1$,
there exists an $(n,k)$-partial agreement algorithm that can tolerate $t$ crash failures,
\begin{enumerate}
\item
\textbf{if} $k \leq \lceil n/ \min(m,t+1) \rceil$.
Furthermore, when $m\leq t+1$ there exists
such an algorithm in which the processes do not need to communicate;
\item
\textbf{only if} $k \leq \lfloor n/ \min(m,t+1) \rfloor + (n \bmod \min(m,t+1))$.
\end{enumerate}
\end{theorem}
A very interesting special case is when $n$ is divisible by $\min(m,t+1)$.
\begin{corollary}
\label{coro:remainder}
For $n\geq2$, $m\geq 2$ and $t\geq 1$,
when $n \bmod \min(m,t+1) = 0$,
there exists an $(n,k)$-partial agreement algorithm that can tolerate $t$ crash failures
if and only if $k \leq n/ \min(m,t+1)$.
\end{corollary}

The proof of the \emph{if} direction of Theorem~\ref{thm:MultivalueAgreement} follows from
Lemma~\ref{lemma:ifDirection}.
The proof of the \emph{only if} direction follows from
Lemma~\ref{lemma:onlyifDirection1}.
Informally, the essence of the proof of Theorem~\ref{thm:MultivalueAgreement} is in
showing that the computational power of
an $(n,\lfloor n/ \min(m,t+1) \rfloor + (n \bmod \min(m,t+1))+1)$-agreement algorithm (object)
is at least as strong as
the computational power of an $(n,n, \min(m,t+1)-1)$-partial set agreement algorithm,
in the presence of $\min(m,t+1)-1$ failures.
Thus, since it is impossible to solve $(n,n, \min(m,t+1)-1)$-partial set agreement
in the presence of $\min(m,t+1)-1$ failures,
it is also impossible to solve $(n,\lfloor n/ \min(m,t+1) \rfloor + (n \bmod \min(m,t+1))+1)$-partial agreement.
In the proof the known result \#2,
regarding the impossibility of solving $(n,n,t-1)$-partial set agreement in the presence of
$t-1$ faulty processes and $m\geq t$ possible input values, is used.
%
\begin{lemma}[if direction]
\label{lemma:ifDirection}
For $n\geq2$, $m\geq 2$, and $t\geq 1$,
there exists an $(n,k)$-partial agreement algorithm that can tolerate $t$ crash failures
\textbf{if} $k \leq \lceil n/ \min(m,t+1) \rceil$. Furthermore, when $m\leq t+1$,
an algorithm exists in which the processes do not need to communicate.
\end{lemma}

\proof
Since, in an asynchronous system,
a crashed process cannot be distinguished from a very slow process,
the agreement requirement can be simplified as follows:
``When all the $n$ processes decide, at least $k$ of them decide on the same value.''

For $m\leq t+1$, consider the following
$(n,\lceil n/m \rceil)$-partial agreement algorithm.
Without communication, each process
simply decides on its own input value.
Thus, at least $\lceil n/m \rceil$ of the processes are guaranteed to decide on the same value
in runs where all the $n$ processes decide (i.e., in fault-free runs).
Since there is no communication,
this simple algorithm satisfies both the weak and strong validity requirements, for any number of $t\geq 1$ of failures.

For $m > t+1$, consider the following
$(n,\lceil n/t+1) \rceil)$-partial agreement algorithm.
Each process $p$ writes its input value into a single writer register
(resp.\ sends its input to everybody)
and continuously reads the single-writer registers of the other processes
(resp.\ waits to receive messages from the other processes)
until it knows the inputs of $n-t$ processes, including itself.
Since there are at most $t$ failures, this procedure will always terminate.
Then, $p$ decides on the maximum input value it knows about.
This reduces the number of decision values to $t+1$.
Hence, this algorithm  solves the $(n,\lceil n/t+1\rceil)$-partial agreement problem.
\qed

\begin{lemma}[only if]
\label{lemma:onlyifDirection1}
For $n\geq 2$, $m\geq 2$ and $t\geq 1$,
there exists an $(n,k)$-partial agreement algorithm that tolerates $t$ crash failures
\textbf{only if}
$k \leq \lfloor n/\min(m,t+1) \rfloor + (n \bmod \min(m,t+1))$.
\end{lemma}

\proof
Recall that strong validity is assumed.
We assume to the contrary that
for $n\geq 2$, $m\geq 2$ and $t\geq 1$,
there exists an $(n,k)$-partial agreement algorithm where $k=\lfloor n/\min(m,t+1) \rfloor + (n \bmod \min(m,t+1)) +1$, called $A$, that can tolerate $t$ crash failures,
and shows that this assumption leads to a contradiction.
Clearly algorithm $A$ is correct with the additional restriction that $m\leq t+1$.
Thus, for the rest the proof of Lemma~\ref{lemma:onlyifDirection1}, it is
assumed that $m\leq t+1$, and hence, $m = \min(m,t+1)$.

For a run $\rho$, we denote by $I(\rho)$ the number of
(different) input (proposed) values in $\rho$;
clearly, $1\leq I(\rho) \leq n$.
By the above assumption,
\begin{quote}
for every run $\rho$ of $A$ in which all the $n$ processes decide,
there exists a proposed \hfill (1)\\
value $v$ such that
at least $\lfloor n/m \rfloor + (n \bmod m) +1$ processes decide on $v$ in $\rho$.
\end{quote}
It follows from (1) and the fact that $n= \lfloor n/m \rfloor *m + (n \bmod m)$,
that,
\begin{quote}
for every run $\rho$ of $A$ in which all the $n$ processes decide,
either $I(\rho)< m$ or there \hfill (2)\\
exists a proposed value $u$ such that
at most $\lfloor n/m \rfloor -1$ processes decide on $u$ in $\rho$.
\end{quote}
It also follows from (1), the fact that $n- (\lfloor n/m \rfloor -1) *m > m-1$, and 
the assumption that $m\leq t+1$,  that
\begin{quote}
for every run $\rho$ of $A$ in which at least $n - (m-1)$ processes decide
\hfill (3)\\
(i.e., there are at most $m-1$ failures in $\rho$),
there exist a proposed value $v$\\
such that at least $\lfloor n/m \rfloor$ processes decide on $v$.
\end{quote}
It follows from (2) and (3) that,
\begin{quote}
for every run $\rho$ of $A$ in which at least $n - (m-1)$ processes decide
\hfill (4)\\
(i.e., there are at most $m-1$ failures in $\rho$),
\begin{enumerate}
\item
there exist a proposed value
$v$ such that in $\rho$, and in any extension of $\rho$,
at least $\lfloor n/m \rfloor$ processes decide on $v$, and
\item
either $I(\rho)< m$ or there exist a proposed value
$u$ such that in $\rho$, and in any extension of $\rho$,
at most $\lfloor n/m \rfloor -1$ processes decide on $u$.
\end{enumerate}
\end{quote}
We use $A$ to construct an $(n,n,m-1)$-partial set agreement (i.e., $(m-1)$-set agreement for $n$ processes)
algorithm for $m$ different input values that can tolerate $m-1$ crash failures, called $B$, as follows:
$B$ works in two (asynchronous) phases of computation:
\begin{enumerate}
\item
Phase one: Each process $p$ participates in $A$ and decides on some proposed value denoted $\decision_p(A)$.
\item
Phase two:
Each process owns a single-writer register.
Each process $p$ writes $\decision_p(A)$ into its single-writer register and repeatedly
reads all the $n$ single-writer registers
until it notices the decision values from the first phase of all
the other processes except maybe $m-1$ of them (since $m-1$ processes may fail).
\end{enumerate}
Let us denote by $V(p)$ the multi-set (i.e., with possible repetitions)
of decision values that $p$ noticed in the second phase. Clearly, $n-(m-1)\leq |V(p)|\leq n$.
Next, $p$ considers only the values in $V(p)$ with the largest number of repetitions,
decides on the largest value among these values, and terminates.
For example, if $V(p)=\{1,1,1,2,2,2,3,3\}$ then $p$ decides on the value 2.

By property (4) above, we have that for every run $\rho$ of $B$ in which at least $n - (m-1)$ processes decide,
\begin{enumerate}
\item
for every process $p$,
there exists a proposed value $v$, such that $v \in V(p)$ and appears
at least $\lfloor n/m \rfloor$ times in $V(p)$.
\item
either $I(\rho)< m$ or there exist a proposed value
$u$ such that in $\rho$, and in any extension of $\rho$,
for every process $p$,
$u$ and appears at most $\lfloor n/m \rfloor -1$ times in $V(p)$.
\end{enumerate}
Thus, there exists a proposed value $u$
that no process will decide on in $B$; and for each process $p$ there is a proposed value
(which appears at least $\lfloor n/m \rfloor$ times in $V(p)$)
that $p$ can decide on.
(We notice that, if $v$ is the value that $\lfloor n/m \rfloor + (n \bmod m ) + 1$
processes decide on in some execution then in a prefix of that execution in which $m-1$
processes fail it is \emph{not} required that at least $\lfloor n/m \rfloor$ processes decide on $v$.)
Thus, the processes will decide on at most $m-1$ different proposed values.
This implies that $B$ is an $(n,n,m-1)$-partial set agreement algorithm that can tolerate $m-1$ faulty process,
violating the known result \#2 (stated in the introduction),
regarding the impossibility of solving $(n,n,m-1)$-partial set agreement in the presence of
$m-1$ faulty processes and $m$ possible input values \cite{BG93,HS99,SZ2000}.
\qed

\begin{corollary}
\label{coro:MultiValueAgreementPlus}
For $n\geq 2$, $m\geq 2$ and $t\geq 1$,
there exists an $(n,k)$-partial agreement algorithm that can tolerate $t$ crash failures
\textbf{only if }
$k \leq  \min_{2\leq \ell\leq\min(m,t+1)} \{\lfloor n/\ell \rfloor + (n \bmod \ell)\}$.
\end{corollary}

\proof
By definition,
any algorithm that solves the $(n,k)$-partial agreement problem
in the presence of (up to) $t$ failures when the maximum number
of possible input values is $m$, must also solve
the $(n,k)$-partial agreement problem
in the presence of $t$ failures when the maximum number
of possible input values is strictly less than $m$.
The result follows from the above observation and Lemma~\ref{lemma:onlyifDirection1}.
\qed


\section{Synchronous Partial Agreement}
\label{sec:Synchronous}
In a synchronous message-passing system in which up to $t$ processes may crash,
there is a simple full agreement algorithm with $t+1$ rounds. Furthermore,
it is known that every full agreement algorithm  requires at least $t + 1$ rounds~\cite{AT1999}.
Can we do better for partial agreement? We show that, in some cases, we might be able to reduce the
number of rounds to $t$, while in other cases we can do much better.
All the results in Section~\ref{sec:Synchronous} hold under both the weak and the strong validity requirements.

\begin{theorem}
\label{thm:lowerBound}
For $n\geq2$, $m\geq 2$, $n-2\geq t\geq1$, and $k \geq \lceil (n+t+1)/2 \rceil$,
in a synchronous message-passing system in which up to $t$ processes may crash,
every $(n,k)$-partial agreement algorithm requires at least $t$ rounds.
\end{theorem}
Informally, the essence of the proof is in
showing that if
an $(n,\lceil (n+t+1)/2 \rceil)$-partial agreement algorithm
can be solved in less than $t$ rounds in the presence of $t$ failures,
then an $(n,n)$-partial agreement algorithm
can be solved in less than $t+1$ rounds in the presence of $t$ failures.
Since it is impossible to solve $(n,n)$-partial agreement in less than $t+1$ rounds
in the presence of $t$ failures,
it is also impossible to solve
$(n,\lceil (n+t+1)/2 \rceil)$-partial agreement
in less than $t$ rounds in the presence of $t$ failures.
In the proof, the known result \#3 is used.

\proof
By definition, any algorithm that solves the $(n,k)$-partial agreement problem
in the presence of (up to) $t$ failures when the maximum number
of possible input values is $m$, must also solve
the $(n,k)$-partial agreement problem in the presence of $t$ failures
when $m=2$. So, since we are proving a lower bound,
for the rest of the proof we assume that $m=2$.

We assume to the contrary that
there exists an $(n,k)$-partial agreement algorithm where $k = \lceil (n+t+1)/2 \rceil$,
called $A$, that can be solved in less than $t$ rounds and can tolerate $t$ crash failures,
and shows that this assumption leads to a contradiction.
Obviously, $k = \lceil (n+t+1)/2 \rceil$ implies the same result for
$k \geq \lceil (n+t+1)/2 \rceil$.

By definition, in any (fault-free) run of $A$ in which all the $n$ processes decide,
there must exist a value $v$ such that
the number of processes that decide on $v$ minus the number of processes
that decide on any other value is at least $t+1$.
To see this, observe that since the value $v$ appears at least $\lceil (n+t+1)/2 \rceil$ times,
the other values appear at most $n- (\lceil n+t+1)/2 \rceil$ times.
So, $\lceil (n+t+1)/2 \rceil -  (n- \lceil (n+t+1)/2 \rceil) \geq t+1$.

This implies that in any run in which at most $t$ processes fail,
and all the other processes decide,
there must exist a value $v$ such that
the number of processes that decide on $v$ minus the number of processes
that decide on any other possible value is at least one.
Thus, in any run of $A$ in which \emph{at most} $t$ processes fail, there is a value $v$ such that
a strict majority (i.e., more than half) of the processes decide on $v$.

We use $A$ to construct an $(n,n)$-partial agreement algorithm, called $B$,
that can be solved in $t$ rounds and can tolerate $t$ crash failures.
$B$ works in two phases, the first takes less than $t$ rounds
and the second takes exactly one round.
\begin{enumerate}
\item
Phase one: Each process $p$ participates in $A$ and decides on some value denoted $\decision_p(A)$.
This takes at most $t-1$ rounds. We notice that it is
possible that no process fails during the first phase.
\item
Phase two:
Each process $p$ send the value $\decision_p(A)$ (from the first phase)
to all the other processes.
Then $p$ waits to receive messages from the other processes
until it learns the decision values from the first phase of all
the other processes except maybe $t$ of them (since $t$ processes may fail).
\end{enumerate}
Recall that we assume $m=2$.
As explained above, in the $n-t$ (or more) decision values
from phase one that $p$ knows about (including its own value),
there must be exactly one value $v$ that was decided upon by more than half of the processes.
So, at the end of phase two, $p$ decides on $v$, and terminates.
This completes the description of algorithm $B$.

We prove that for $t\geq 1$, in $B$ all the non-faulty processes decide on the same proposed value $v$.
For a given run, assume that $f$ processes, where $1 \leq f \leq t$ has failed.
Thus $n-f$ processes succeeded in sending their decision values from phase one to
all the other processes. In such a case, as explained above,
there must exist a value $v$ such that (in phase one)
the number of processes that decided on $v$ minus the number of processes
that decided on any other possible value is at least $t-f+1 \geq 1$.
Thus, since each of the non-faulty processes see (at the end of phase two)
a subset of size at least $n-t$ values, they will all decide on the same value $v$.

Thus, $B$ is an $(n,n)$-partial agreement algorithm that requires less than $t+1$ rounds
in the presence of $t$ failures. However, this violates
the known result \#3 (as stated in the introduction),
regarding the impossibility of solving $(n,n)$-partial agreement
in less than $t+1$ rounds in the presence of $t$ faulty processes \cite{AT1999}.
\qed

\noindent
Next, we observe that it is possible to significantly reduce the number of rounds in some cases.
This simple observation follows easily
from a known result regarding the number of rounds that are sufficient
for solving the set agreement problem \cite{AW98,CHLT2000}.

\begin{theorem}
\label{thm:upperBound}
For $n\geq2$, $m\geq 2$, $t\geq1$, and $\ell \geq 1$,
in a synchronous message-passing system in which up to $t$ processes may crash,
there exists an $(n,\lceil n/\ell \rceil)$-partial agreement algorithm with $\lfloor t/\ell\rfloor +1$ rounds.
\end{theorem}

\proof
A simple algorithm was presented in \cite{AW98,CHLT2000} that solves $\ell$-set agreement
and requires only $\lfloor t/\ell\rfloor +1$ rounds. This algorithm clearly solves also $(n,\lceil n/\ell \rceil)$-partial agreement.
For completeness, we give below a description of the algorithm and an explanation.
\begin{quote}
\textit{The algorithm}~\cite{AW98,CHLT2000}:
The algorithm consists of exactly $\lfloor t/\ell\rfloor +1$ rounds.
In each round, every process sends a message with its preferred value (initially its input)
to all the processes (including herself)
and waits until the end of the round
to receive all the messages that were sent to it during the round.
From the set of all messages that the process has received
in a given round,
it chooses the minimum value as its new preferred value.
Then, it
continues to the next round supporting this (minimum) value as its new preferred value.
The algorithm terminates after $\lfloor t/\ell\rfloor +1$ rounds, and each process decides on its
preferred value at the end of the last round.
\end{quote}
Explanation. If at some round, $x$ processes fail, then in the next round
the number of different values will be \emph{at most} $x+1$, and this number will never
be increased in subsequent rounds. Thus, the worst case is when the number
of faults is the same in each round.
Given $t$ faults and $\lfloor t/\ell\rfloor +1$ rounds, it is only possible to arrange  that
in \emph{each} round there will be at least
$$x = \left\lfloor \frac{t}{\lfloor t/\ell\rfloor +1} \right\rfloor = \ell -1$$ faults in each round.
Thus, the maximum number of \emph{different} decision values that is possible is $x+1 =\ell$.
\qed

\section{Partial Agreement Using Strong Shared Objects}
\label{sec:shrongSM}
The results presented in this are for the $SM(g)$ computational model.
Recall that wait-free full agreement objects for $g$ processes
are shared objects that solve the full agreement problem for $g$ processes
in the presence of any number of failures \cite{Her91}.
All the results in Section~\ref{sec:shrongSM} hold under both the weak and the strong validity requirements.

\begin{theorem}
\label{Thm:strongObjects}
For $n\geq 2$, $m \geq 2$ and $n>t\geq 1$,
there exists an $(n,k)$-partial agreement algorithm that can tolerate $t$ crash failures
using atomic registers and wait-free full agreement objects for $t$ processes,
only if $k \leq \lceil (n+t-1)/2 \rceil$.
\end{theorem}
The proof is an adaptation of the one used for proving Theorem~\ref{Thm:BinaryAgreement}.
Informally, the essence of the proof is in
showing that an $(n,\lceil (n+t-1)/2 \rceil+1)$-partial agreement algorithm (object)
has the same computational power as an $(n,n)$-partial agreement algorithm,
in the presence of $t$ failures, in the $SM(t)$ model.
Thus, since it is impossible to solve $(n,n)$-partial agreement,
it is also impossible to solve $(n,\lceil (n+t-1)/2 \rceil+1)$-partial agreement.
In the proof, the known result \#4, is used.

\proof
By definition, any algorithm that solves the $(n,k)$-partial agreement problem
in the presence of (up to) $t$ failures when the maximum number
of possible input values is $m$, must also solve
the $(n,k)$-partial agreement problem in the presence of $t$ failures
when $m=2$. So, 
for the rest of the proof we assume that $m=2$.

We assume to the contrary that
there exists an $(n,k)$-partial agreement algorithm where $k=\lceil (n+t-1)/2 \rceil +1$, called $A$,
that can tolerate $t$ crash failures, and shows that this assumption leads to a contradiction.

By definition, in any (fault-free) run of $A$ in which all the $n$ processes decide,
there must exist a proposed value $v$ such that
the number of processes that decide on $v$ minus the number of processes
that decide on any other possible value is at least $t+1$.
To see this, observe that since the value $v$ appears at least $\lceil (n+t-1)/2 \rceil +1$ times,
the other possible values appear at most $n- (\lceil (n+t-1)/2 \rceil +1)$ times.
So, $(\lceil (n+t-1)/2 \rceil +1) -  (n- (\lceil (n+t-1)/2 \rceil +1)) \geq t+1$.

This implies that, in any run in which at most $t$ processes fail,
and all the other processes decide,
there must exist a proposed value $v$ such that
the number of processes that decide on $v$ minus the number of processes
that decide on the other possible value is at least one.
Thus, in any run of $A$ in which at most $t$ processes fail, there is a proposed value $v$ such that
a strict majority of the processes decide on $v$.

We use $A$ to construct an $(n,n)$-partial agreement algorithm that can tolerate $t$
crash failures, called $B$, as follows:
$B$ works in two (asynchronous) phases of computation.%
A process that finishes phase one,
immediately starts participating in phase two.
\begin{enumerate}
\item
Phase one: Each process $p$ participates in $A$ and decides on some proposed value denoted $\decision_p(A)$.
\item
Phase two:
Each process $p$ owns a single-writer register, and
initially writes $\decision_p(A)$ in a single-writer register.
Then $p$ repeatedly
reads the single-writer registers of the other processes
until it learns the decision values from the first phase of all
the other processes except maybe $t$ of them (since $t$ processes may fail).
\end{enumerate}
Recall that $m=2$.
As explained above, in the $n-t$ decision values
from phase one that $p$ knows about (including its own value),
there must be one proposed value $v$ that was decided upon by more than half of the processes.
So, at the end of phase two, $p$ decides on that value $v$, and terminates.
This completes the description of algorithm $B$.
We prove that when $t\geq 1$, in $B$ all the non-faulty processes decide on the same value $v$.
For a given run, assume that $f$ processes, where $1 \leq f \leq t$ has failed.
Thus $n-f$ processes succeeded in writing their decision values from phase one into their
single-writer registers. In such a case, as explained above,
there must exist a value $v$ such that (in phase one)
the number of processes that decided on $v$ minus the number of processes
that decided on the other possible value is at least $t-f+1 \geq 1$.
Thus, since each the non-faulty processes see (at the end of phase two)
a subset of size at least $n-t$ values, they will all decide on the same proposed value $v$.

Thus, $B$ is an $(n,n)$-partial agreement algorithm that can tolerate $t$ faulty process,
violating the known result \#4 (as stated in the introduction),
regarding the impossibility of solving $(n,n)$-partial agreement
in the presence of a $t$ faulty process using
using atomic registers and wait-free full agreement objects for $t$ processes
\cite{Her91}.
\qed

\noindent
Next, we demonstrate that in the $SM(g)$ model, when $g\geq 2$,
we can do better than the sufficient condition presented in Theorem~\ref{Thm:BinaryAgreement}.

\begin{theorem}
\label{thm:alg:StrongSM}
For $n\geq 4$, $m\geq 2$ and $t\geq 1$,
there exists an $(n,k)$-partial agreement algorithm that can tolerate $t$ crash failures
using atomic registers and wait-free full agreement objects for $g$ processes
if $k \leq \max(g, 3\lfloor \min(\lfloor n/2 \rfloor ,g)/2 \rfloor )$.
\end{theorem}

\proof
We consider two cases.
The first case is when $g > 3\lfloor n/4 \rfloor$.
To solve this case, a single wait-free full agreement object for $g$ processes,
called $A$, is used. We let $g$ processes participate in $A$ and
decide on the same value. Each one of the other $n-g$ processes simply decides on its input.

The second case is when $g \leq 3\lfloor n/4 \rfloor$.
Let $\hat{g}=\min(\lfloor n/2 \rfloor ,g)$.
Three wait-free full agreement objects for $g$ processes,
called $A$, $B$ and $C$, are used for the algorithm.
Two disjoint groups of processes, $G_1$ and $G_2$ are created
each of size $\hat{g}$.
Then, we choose $\lfloor \hat{g}/2 \rfloor$ processes from $G_1$ and
$\lfloor \hat{g}/2 \rfloor$ processes from $G_2$, and
create the groups $G_3$ and $G_4$, respectively.
Each process that does not belong to $G_1$ or $G_2$ simply decides on its input and terminates.
The processes in $G_1$ first participate in $A$, and
then the processes in $G_3$ use the decision value from $A$ as their
new input, participate in $C$ and use the decision value from $C$ as their
final decision value. The other processes in $G_1$ use the decision value from $A$
as their final decision value.
Similarly,
The processes in $G_2$ first participate in $B$, and
then the processes in $G_4$ use the decision value from $B$ as their
new input, participate in $C$ and use the decision value from $C$ as their
final decision value. The other processes in $G_2$ use the decision value in $A$
as their final decision value. The result follows.
\qed\\
\\
We notice that,
by Theorems~\ref{Thm:strongObjects} and \ref{thm:alg:StrongSM},
for the $SM(n/2)$ model, the bound is tight when $n=t/2$ and $n$ is divisible by 4.
\begin{corollary}
\label{coro:strongObjects}
Assume that $m \geq 2$ and $n$ is divisible by 4.
There exists an $(n,k)$-partial agreement algorithm that can tolerate $n/2$ crash failures
using atomic registers and wait-free full agreement objects for $n/2$ processes,
if only if $k \leq 3n/4$.
\end{corollary}
%

\section{Related Work and a few open problems}


The consensus (full agreement) problem is a fundamental coordination problem and is at the
core of many algorithms for distributed applications.
The problem was formally presented in \cite{PSL1980,LSP1982}, in which Byzantine failures are assumed.
In Byzantine (malicious) failures -- the strongest type of failures -- there are
no restrictions on how a process may operate once it fails.
Achieving (full) Byzantine agreement require $\Omega(t)$
connectivity in the communication network in
order to tolerate $t$ Byzantine faults \cite{Dolev1982}.
A simple corollary of this result is that a system can reach
agreement in the presence of $t$ faulty processes,
only if every process (processor) is directly connected
to at least $O(t)$ others.

Motivated by the need to achieve Byzantine agreement
on sparse networks, the notion of \emph{almost everywhere agreement}
was introduced in \cite{DPPU1988}, in
which all but a small number of the correct
processes must choose a common decision
value. Intuitively, the correctness condition is
relaxed by ``giving up for lost'' those correct processes whose communication paths
to the remainder of the network are excessively corrupted by faulty processes.
As an intermediate step in defining almost everywhere agreement, the notion
of $t$-resilient $X$ agreement was introduced, in which (1) when at most $t$ processes
fail all but $X$ of the correct processes must eventually decide on a common value, and
(2) ``if all correct processes begin with the same value $v$, then $v$ must be the common decision value'' \cite{DPPU1988}.
Condition \#1 above is similar to our agreement requirement (in the definition of (\emph{n,k})-partial agreement);
condition \#2 is much weaker than ours weak validity requirement.

It was shown in \cite{DPPU1988} that \emph{synchronous} almost everywhere (Byzantine) agreement can be
achieved even on bounded degree networks,
as long as the number of faults is bounded
by $O(n/ log~n)$, where $n$ is the size of the network.
Later it was shown in \cite{Upfal1992} that such agreement
is also achievable in the presence of up to $O(n)$ faulty processes.

Another related definition was introduced in \cite{Tau2018journal},
in which the traditional notion of fault tolerance is generalized
by allowing a limited number of participating correct processes not to terminate in the presence of faults.
Every process that does terminate is required to return a correct result.
For this generalization, various results regarding the solvability of problems like
election, consensus and renaming using atomic registers are presented.

The most recent related definition is that of bounded disagreement,
which limits the number of processes that decide differently from
the plurality \cite{CHS2020}. Here the agreement requirement is similar
to that of $X$ agreement (and hence also ours); the validity
requirement used is strong validity.
The main result of \cite{CHS2020} is that
there are infinitely many instances of the bounded disagreement task that are not equivalent
to any consensus task and any set agreement task.
None of our results overlaps or can be derived from
the results in \cite{DPPU1988,Tau2018journal,CHS2020}.

Many (deterministic and randomized) consensus algorithms have been proposed for shared memory systems.
Few examples are
\cite{Abrahamson1988,AAC2008,ASS2002,aspnes1990fast,FMT93,LA87,MRR2003,Plo89,RT2020,SSW91,Tau2017podc}.
Dozens of papers have been published on solving the consensus problem in various
message passing models. A few examples are \cite{AbMS2019,DDS87,DLS88,Fi83,FLM86,FLP85,KTZ2021,Maynal2010,YMRGA2019}.
For a survey on asynchronous randomized consensus algorithms, see \cite{Aspnes2003}.
A challenging research direction is to explore
when it is possible to design (deterministic and randomized) partial agreement
algorithms with better time, space, or message complexities than those of
known full agreement algorithms.

The impossibility result that no full agreement algorithm
can tolerate a single crash failure in an asynchronous message-passing system
was proven in \cite{FLP85}.
Proof of an impossibility result for solving
agreement in an asynchronous shared memory system that supports only
atomic read/write registers in the presence of a single crash failure
appeared in \cite{LA87}.
%

The impossibility result for full agreement is a special case
of an impossibility result for the set agreement problem.
The set agreement problem was defined in \cite{Chaudhuri1993};
the impossibility result was proved in \cite{BG93,HS99,SZ2000}.
Many related impossibility results can be found in \cite{AE2014}.
The set agreement problem is a special case of
the partial set agreement problem (defined in the introduction), and it would be interesting
to find out if the impossibility result for set agreement can be
extended to cover more cases of the partial set agreement problem.

Recently it was shown  in \cite{AAEGZ2019,ACR2021,BE2021}
that, for asynchronous SM or MP systems, the proof technique used to prove
the impossibility result for full agreement,
called \emph{extension-based proofs}
(a result which we rely on for proving Theorem~\ref{Thm:BinaryAgreement}),
is not applicable for proving the
impossibility results for set agreement (a result which we rely on for proving Theorem~\ref{thm:MultivalueAgreement}).
It would be interesting to find out if it is also impossible to use
extension-based proofs to prove Theorem~\ref{thm:MultivalueAgreement}.
Also, the necessary condition proved for the synchronous case
(i.e., Theorem \ref{thm:lowerBound}) uses extension-based proofs (i.e., is based on \cite{AT1999});
is it possible to generalize this necessary condition using extension-based proofs only?

Three extensively studied progress conditions
are wait-freedom \cite{Her91}, non-blocking \cite{HW90}
and obstruction-freedom \cite{HLM2003}.
Recently, it has been shown in \cite{EGZ2018}, that
any obstruction-free full consensus algorithm for $n$ processes
using registers must use at least $n$ registers.
It would be interesting to find whether this space lower bound
holds for obstruction-free partial consensus algorithms when
communication is necessary.
Also, it would be interesting to know when wait-free partial agreement
is solvable using various known shared objects which are
stronger than atomic register.

A simple and elegant proof
that in a synchronous system with up to $t$ crash failures
solving agreement requires at least $t + 1$ rounds appeared in \cite{AT1999}.
This result for crash failures is stronger than a similar earlier result
for Byzantine failures ~\cite{FL1982}.
In \cite{CHLT2000}, a tight bound of $\lfloor t/k \rfloor +1$ rounds, is presented,
for solving $k$-set agreement in a synchronous system with up to $t$ crash failures.
The lower bound proof of this result is the first to apply topological
techniques to the synchronous model.

As already mentioned, the Byzantine agreement and Byzantine fault tolerant (BFT) in general
originated from two seminal papers \cite{PSL1980,LSP1982}.
These two papers contain upper and lower bounds for the case where
more than \emph{two-thirds} of the processes are correct.
It is interesting to note that when assuming unforgeable signed messages,
the problem is solvable for any number of Byzantine failures \cite{LSP1982}.
BFT has been intensively investigated for over 40 years now.
Is it possible to design Byzantine partial agreement algorithms with
better time, space, and message complexities than those of known
Byzantine full agreement algorithms?

Algorithms for synchronous systems,
in which communication is done in ``rounds,'' are not practical.
Algorithms for asynchronous systems, where no assumption is made about
the relative speed of the participating processes, are practical and  operate properly in any system.
However, this comes at the cost of efficiency and sometimes even solvability.

One solution is to design \emph{indulgent} algorithms.
The appeal of indulgent algorithms lies in the
fact that when they are executed in an asynchronous system,
they ``lie in wait'' for a short period of time during
which certain timing constraints are met, and when this happens,
these algorithms take advantage of the situation and
efficiently complete their mission \cite{G00,GR04}.
The most known indulgent full agreement algorithm
is the Paxos algorithm \cite{Lamport1998}.
A more recent indulgent agreement algorithm is HotStuff \cite{YMRGA2019}.
An interesting question is whether it is possible to increase efficiency by
designing an  indulgent partial agreement algorithm.

\section{Discussion}
The fundamental full agreement problem has been intensively studied for over 40 years.
It is intriguing to revisit some of the numerous questions and results for this problem
and study them in the context of the partial agreement problem.
For example, when is it possible to design randomized partial agreement algorithms with
better time, space, and message complexities than those of known randomized full agreement algorithms?
Similarly, what are the time and space complexities of obstruction-free partial agreement algorithms?
How many Byzantine failures can partial agreement algorithms tolerate,
and what is the complexity of such Byzantine partial agreement algorithms?
Studying the partial set agreement problem
defined in the introduction would be interesting.
Apart from presenting the problem and the new technical results,
the significance of the article is in exposing open problems
that hopefully will stimulate further research on the partial agreement problem.


\bibliography{../REF/mydatabase,../REF/list,../REF/newRefAll2012,../REF/BibBiology,../REF/newRefAll2022}

\end{document}